# A Comparison of Radial Intensity Profiles of Termination Shock Particles and Anomalous Cosmic Rays in the Outer North-South Heliosheaths Using CRS data from V1 and V2

W.R. Webber[1] and D.S. Intriligator[2]

1. New Mexico State University, Department of Astronomy, Las Cruces, NM, 88003, USA

2. Carmel Research Center, Space Plasma Laboratory, Santa Monica, CA, 90406, USA

# ABSTRACT

Using data from the Voyager 1 and 2 CRS telescopes available on the web through the end of 2014 we have studied the intensity variations of termination shock H nuclei and anomalous cosmic ray H and He nuclei as a function of radial distance. In contrast to the inner part of the heliosheath where the intensity vs. radius profiles in the North and South heliosheaths are much different, these intensity vs. radius profiles, as well as the intensities themselves, are more similar in the outer North and South heliosheaths as measured by V1 and V2 respectively. In the N heliosheath, taken to be 27.6 AU thick beyond the HTS crossing distance of 94 AU, the intensities of termination shock particles and anomalous cosmic rays reach a maximum at between 110-112 AU or at a location ~halfway between the termination shock and the heliopause. They then decrease more or less continuously to an intensity ~2/3 of the maximum for each component just before the final dropout at ~121 AU, just inside the heliopause. This intensity-time behavior favors theories in which the acceleration and loss effects lead to a maximum near the middle of the heliosheath.

The intensity-time profiles and the intensities in the outer part of the South heliosheath observed by V2 are very similar to those in the North observed by V1, with the intensity maxima observed between 96-98 AU, 12-14 AU beyond the helioshock crossing distance of 83.7 AU. At the end of 2014 when V2 was at 107 AU, the intensities of all three components have also decreased to ~2/3 of the maximum intensity in the S heliosheath, the same as they were for V1 just before the heliopause crossing.

Considering several corresponding features between the North and South heliosheaths we estimate a South heliopause crossing by V2 will occur between 2015.0 and 2015.5 when V2 is between ~107 to 108.5 AU from the Sun. The ratio of this distance to the N heliopause distance of 121.7 AU would by 0.89 which is the same as the ratio of the South/North helioshock crossing distances.

**Introduction**

Voyager 1 (V1) crossed the heliopause (HP) at a distance of 121.7 AU from the Sun, approximately 7.7 years after crossing the heliospheric termination shock (HTS) at a distance of 94.1 AU (Stone, et al., 2013). During this time V1 sampled the N heliosheath of thickness ~27.6 AU at a latitude of between 28-32° in the N hemisphere.

Voyager 2 (V2) crossed the HTS at a distance of 83.7 AU and now at the end of 2014, about 7.5 years later, is at ~107 AU in the S hemisphere at a latitude ~30°. During this time V2 has sampled the S heliosheath over a distance of ~23.3 AU beyond the HTS. If the N and S heliopause distances scale in the same way as the N-S helioshock distances, then a S heliopause will be encountered at about 108.2 AU corresponding to a thickness of 24.5 AU.

With these facts as a background we will examine the intensity time profiles of termination shock particles (TSP), generally described as protons of a few MeV, and anomalous cosmic rays, generally described as protons, helium and oxygen nuclei of energies greater than several MeV (see e.g., Cummings and Stone, 2012) in the outer N and S heliosheaths.

In contrast to the situation in the inner heliosheaths, within 10-12 AU of the HTS, where the intensities and intensity time profiles in the N and S heliosheaths are in fact quite different; in the outer heliosheaths the intensities are more comparable as they increase in both hemispheres to a maximum about 12-15 AU beyond the HTS and the intensity time profiles show several possible corresponding N and S features that enable one to relate these features in both hemispheres.

The connection between the two classes of particles, TSPs and ACRs, is not yet fully clear. This uncertainty is also true for the acceleration of both types of particles, especially anomalous cosmic rays. There are theories that propose to accelerate these particles near the HTS, near the HP, and everywhere in between. If we try to provide a list of references here we will surely offend somebody. Suffice to say here is that this acceleration process is a major drain on the solar wind energy that dominates the inner parts of the heliosphere.

Comparing the N-S radial intensity profiles will provide clues as to the location of this acceleration process, and the relative intensities in the N-S heliosheaths, as well as the relative intensities of TSPs vs. ACRs and protons vs. helium nuclei, will provide clues related to the acceleration process itself.

**Data**

The data, which comes from the CRS web-site (http://voyager.gsfc.nasa.gov) is shown in three figures, each of which shows both the V1 and V2 intensities as a function of time, but with V1 delayed by 2.7 years, the time difference between HTS crossings of V1 and V2. The fact that V1 is moving outward faster than V2 by about the same factor (1.125) that the V1 HTS distance is larger than V2 means that events in the N and S heliosheath will occur at nearly the same times at V1 and V2 in this representation (and therefore relative locations in the heliosheath) if the N-S heliopause distances scale the same as the N-S HTS distances.

In Figure 1 we show the 0.5 MeV intensities observed at V1 and V2 in the outer heliosheath. These intensities represent mainly MeV protons and are a proxy for TSPs.

In Figure 2, protons from 6-42 MeV at V1 and V2 are shown. These intensities represent ACR-H of energies ~10 MeV.

In Figure 3, helium nuclei from 6-42 MeV/nuc at V1 and V2 are shown. These intensities represent ACR-He of energies ~10 MeV/nuc.

It should be noted that the intensity-time curves in all three figures are much smoother in the N heliosheath at V1, than at V2 in the S heliosheath.

No normalization between V1 and V2 is made. Certain features that are common to each of the figures are indicated. This includes ① (the point labeled "1" in the figures), the time of maximum at each energy and for each species.

There is also an intensity peak ② (the point labeled "2" in the figures), seen for all species but most clearly in the >0.5 rate (TSPs). At V1 this peak occurred at a distance ~3.0 AU inside the HP and was followed by a more or less continuous decrease in intensity of TSPs until the final dropout at the time of the HP crossing. A dashed line connecting the intensities at the peak and those before the final dropout for each spacecraft and for each species is also shown.

At V2 in the S hemisphere there are also two prominent decreases, shown as shaded regions in Figures 1 and 2, with minima at 2013.02 and 2013.92, which are not seen at V1. There are also, at V2, several intensity spikes of duration a few days, indicated by circles in Figure 1, which are not seen at V1.

## Discussion and Comparison of the Intensities and Variations Seen in the N-S Heliosheaths

The time ① is the time of peak intensity seen in all figures in both of the N-S heliosheaths. In the N heliosheath this time is roughly the same in all figures, the same for TSPs and ACRs, H or He. It occurs at between 0.55-0.60 of the distance between the HTS and the HP, near the middle of the N heliosheath. We believe that this location, as well as the relatively smooth, rising and then decreasing, intensity time profiles, especially in the N heliosheath, favors models in which the acceleration occurs throughout the N heliosheath with a peak efficiency somewhere near the center of the heliosheath.

The peak ① at V2 also occurs at roughly the same time for both TSPs and ACRs. The relative intensities of TSPs and ACRs at the times of peak intensities are also the same in each heliosheath to within $\pm$ 10%. The time of the peak intensities ① in the S heliosheath is 0.7-0.9 of a year "earlier" in our method of plotting in all cases. This could suggest that the S heliosheath is further squashed in thickness relative to the N heliosheath in comparison to that expected from the ratio of 1.125 for the N-S HTS crossing distances.

At V2 for ACR-He there are two after-peaks. These peaks, occurring at about 2011.95 and 2012.54, are correlated with two prominent pressure jumps seen in the solar wind plasma data on the MIT data site, http://web.mit.edu/afs/athena/org/s/space/www/voyager/voyager_data /voyager_data.html. The intensities at the times of these after-peaks are ~10% higher than the initial peak and ~30% higher than the maximum intensity of ACR-He that is observed for these particles by V1 in the N heliosheath.

The intensity peak we call ② occurs later at V1 when this spacecraft was only about 3 AU inside the heliopause. After this peak, which is observed in varying degrees for TSPs and ACR H and He nuclei at V1, the intensities of all particles decreases continuously until the time of the heliopause crossing. The time of the peak intensity (at ~2011.8 at V1) is a time of great variability in the magnetic field. A maximum of ~0.3 nT and a minimum ~0.07 nT are observed between 2011.7 and 2011.9 (see Burlaga and Ness, 2013).

We believe that a corresponding intensity peak ② is observed at about 2014.2 at V2. It is seen at varying levels in all components, but weaker for ACRs, the same as in the N heliosheath by V1. This "event" ② at V2 again occurs earlier than in the N heliosheath, this time between 0.4-0.6 years earlier. The intensities of all components at V2 have also continued to decrease irregularly up to 2015.0, as they did in the corresponding time interval at V1 as it approached the HP.

If these are indeed the correct N-S heliosheath correspondences that we have used, V2 could cross the S heliopause as early as 2015.0 at a distance of about 107 AU from the Sun. This would give a S heliosheath thickness of 23.3 AU. Recall that if we assume the S heliosheath thickness is squashed by the same fraction as the S/N ratio of the HTS crossing distances, the S heliosheath thickness would be at ~24.3 AU and the S heliopause crossing by V2 would then occur about the middle of 2015.

We also note from Figures 1, 2 and 3 the general <u>decrease</u> in intensity of TSPs and ACRs at V1 from the time of maximum to the time of the HP crossing. In fact, just before the HP crossing, the intensities of all of these components are at their minimum in the outer heliosheath, about 0.7 times the intensities at maximum, an important feature for theoretical models.

The rate of decrease of intensity after the maximum ① at each spacecraft is shown in Figures 1, 2 and 3 as a simple exponential line for guidance purposes. These curves are extended to the time just before the final decrease at V1 associated with the HP and to the end of data at V2. This dashed line illustrates the overall intensity decrease better for the ACRs than for the TSPs which have many localized variations in the S heliosheath as discussed below.

Of particular note are the two large decreases seen in TSPs in Figure 1 at V2. These decreases are also seen more weakly in ACR-H at V2. They are not seen at V1. At V2 the intensity minima for TSP's occur at 2013.02 and 2013.92, about a year apart. We believe that these decreases may be related to the outward propagation of major shocks through the S heliosheath and passing the approximate location of V2 at these specific times. In fact, Liu, et al., 2014, and Intriligator, et al., 2014, have calculated that the shock that produced the radio emission seen by V1 between days 80-90 of 2013, when it was beyond the HP (Gurnett, et al., 2014), would have reached the distance of V2 in late 2012 or early 2013, the time we observe the first minimum. This shock originated on the Sun in March 2012, and would have had a propagation time to V1 ~400 days. We will discuss more about both of these decreases and their relation to outward propagating shocks in a separate paper, (Intriligator and Webber, 2015).

Another unusual feature seen in Figure 1 for TSP in the S heliosheath by V2 but not in the N heliosheath, are sudden intensity spikes of a few days duration. Depending upon how one defines them, there are 6-8 of these spikes as noted in Figure 1 with small circles. Upon close examination of the MIT plasma data on the web, we find, in every case, a sudden increase in plasma pressure, occurring within $\pm$ 1 week of each sudden TSP increase (the plasma pressure is a 7 day running average). These increases are not so evident in the plasma speed profiles and may be related to plasma density increases.

## Summary and Conclusions

We present radial intensity profiles of TSPs and ACRs observed by V1 and V2 in the N and S outer heliosheaths. These particles are believed to be accelerated in the heliosheath at its inner and outer boundaries or in between but there is great debate as to where and how this acceleration occurs.

We find that the maximum intensity of both TSPs and ACR H and He nuclei in both the N and S heliosheaths (V1 and V2) occurs at a distance about 0.50-0.60 times the distance between the HTS inner boundary and HP outer boundary, that is about halfway through the heliosheath. In the N hemisphere V1 sees very smooth radial intensity profile for both ACR H and He nuclei, with a decreasing intensity as V1 approaches the HP. Just inside the HP, but before the final intensity drop out, these intensities are about 2/3 of their maximum intensities. The radial profile for TSPs in the S heliosheath is similar but with a lot more variability.

However, the intensity at the end of 2014, is also about 2/3 of the intensity at the peak over three years earlier.

The intensity vs. time (radial) curves for each component are more variable and structured at V2 in the S hemisphere. The intensities all rise to a sharp peak in the S heliosheath which is within 10% of that observed in the N hemisphere for all components. The peak in the S heliosheath occurs 0.6-0.8 year earlier than in the N heliosheath on the relative timescale used here.

The radial intensity curves in both hemispheres, which have maximum at a location between 0.55-0.60 of the distance between the HTS and HP, make a strong case for acceleration throughout the heliosheath with a maximum efficiency in the middle of the heliosheath radially.

Large intensity decreases at V2 in the S hemisphere, most prominent for the TSPs but also seen for ACR-H, with minima at 2013.02 and 2013.92, are not seen at V1. We believe that these decreases are related to the outward passage past the V2 location, of large solar originated shocks. In fact, the time of the first minimum at 2013.02 is consistent with the passage of a shock (Liu, et al., 2014; Intriligator, et al., 2014) that was later determined to be the source of radio emission seen at V1 beyond the HP at about 2013.22 (Gurnett, et al., 2013).

Several sharp spikes of TSP intensity are also seen at V2. These spikes of a few days duration are related in all cases to sudden increases in plasma pressure observed on V2.

And finally, the events designated ① and ② at V1 and V2 appear to occur "earlier" at V2. This could indicate that the S heliosheath is squashed in thickness relative to N heliosheath. Nominally, if both heliosheaths have the same relative thicknesses as the inner heliosphere as determined from the HTS crossing distances, then the S heliosheath would be 24.5 AU thick (compared to 27.6 AU in the N). As a result, the S heliopause would be at ~108 AU and would be crossed by V2 near the middle of 2015. For a squashed heliosheath, as implied by the relative time of corresponding N and S heliosheath events, ① and ②, this crossing could occur as early as 2015.0.

**Acknowledgements:** The data used here comes from the CRS public website (http://voyager.gsfc.nasa.gov/heliopause/data.html) carefully maintained up to date by Nand Lal. WRW greatly appreciates the support of his other CRS colleagues, PI Ed Stone and Alan Cummings and Bryant Heikkila, as well as JPL for their financial support. DSI was supported by Carmel Research Center, Inc.

**Figure Captions**

**Figure 1:** Intensity time variations of >0.5 MeV particles (e.g., TSPs) in the outer heliosheaths at V1 in the N and V2 in the S. The V1 time is delayed 2.7 years to account for the time differences in crossing the helioshock. The labels ① and ②, the dashed lines and the small circles are described in the text. The shaded regions, which show major decreases in TSPs at V2 in the S heliosheath, are also described in the text.

**Figure 2:** Similar to Figure 1, but for 6-42 MeV H nuclei which are mainly ACRs.

**Figure 3:** Similar to Figure 1, but for 6-42 MeV/nuc He nuclei which are mainly ACRs.

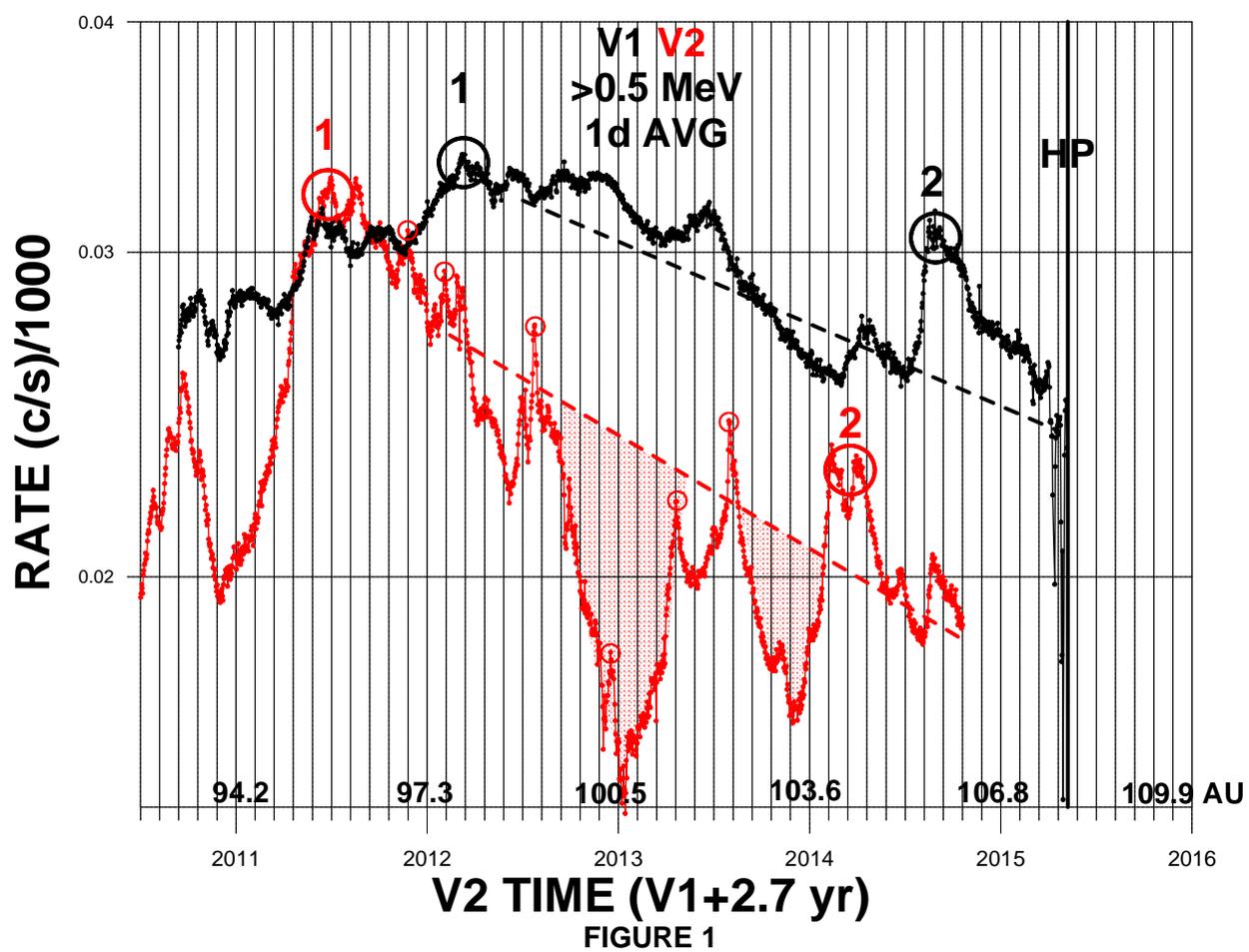

FIGURE 1

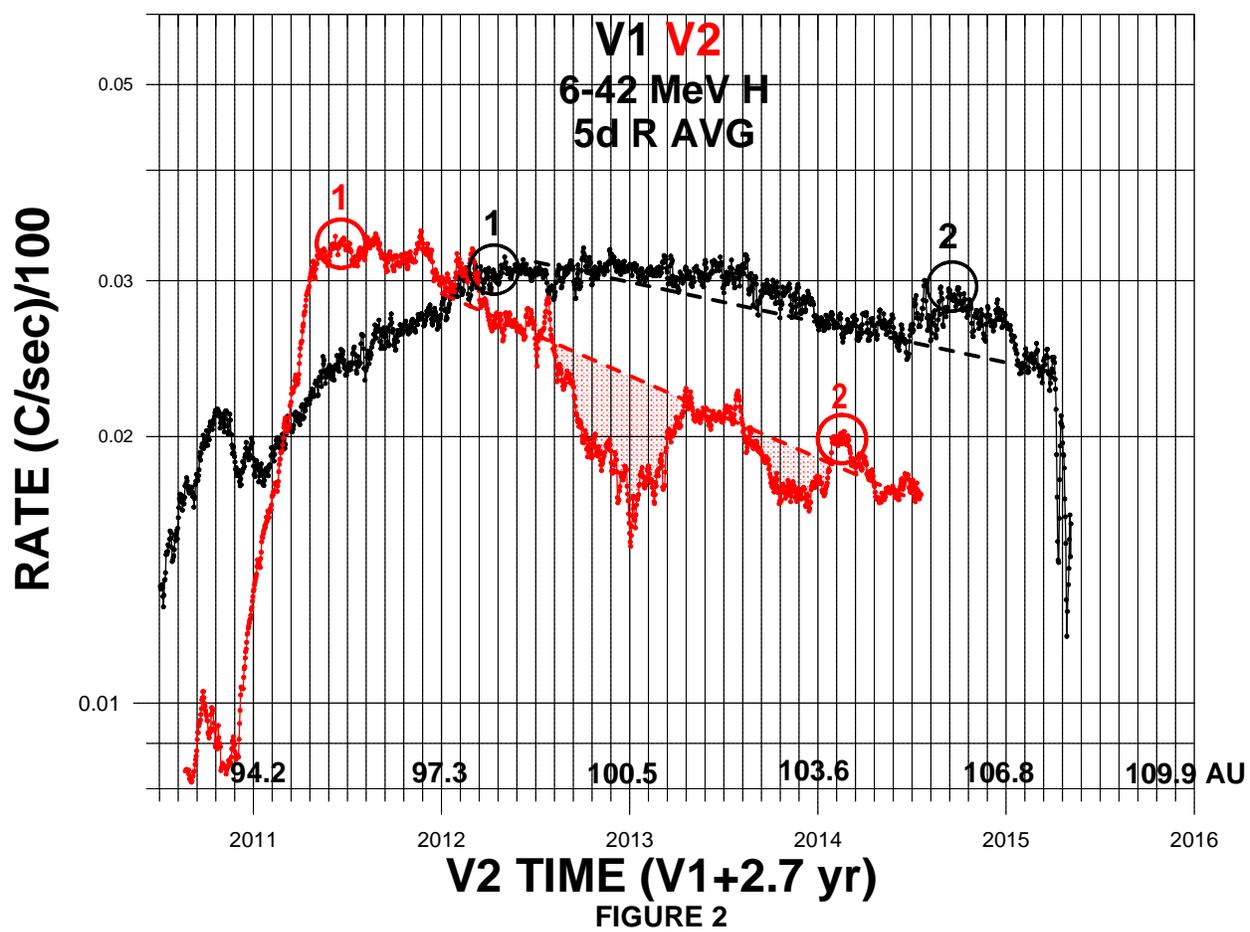

FIGURE 2

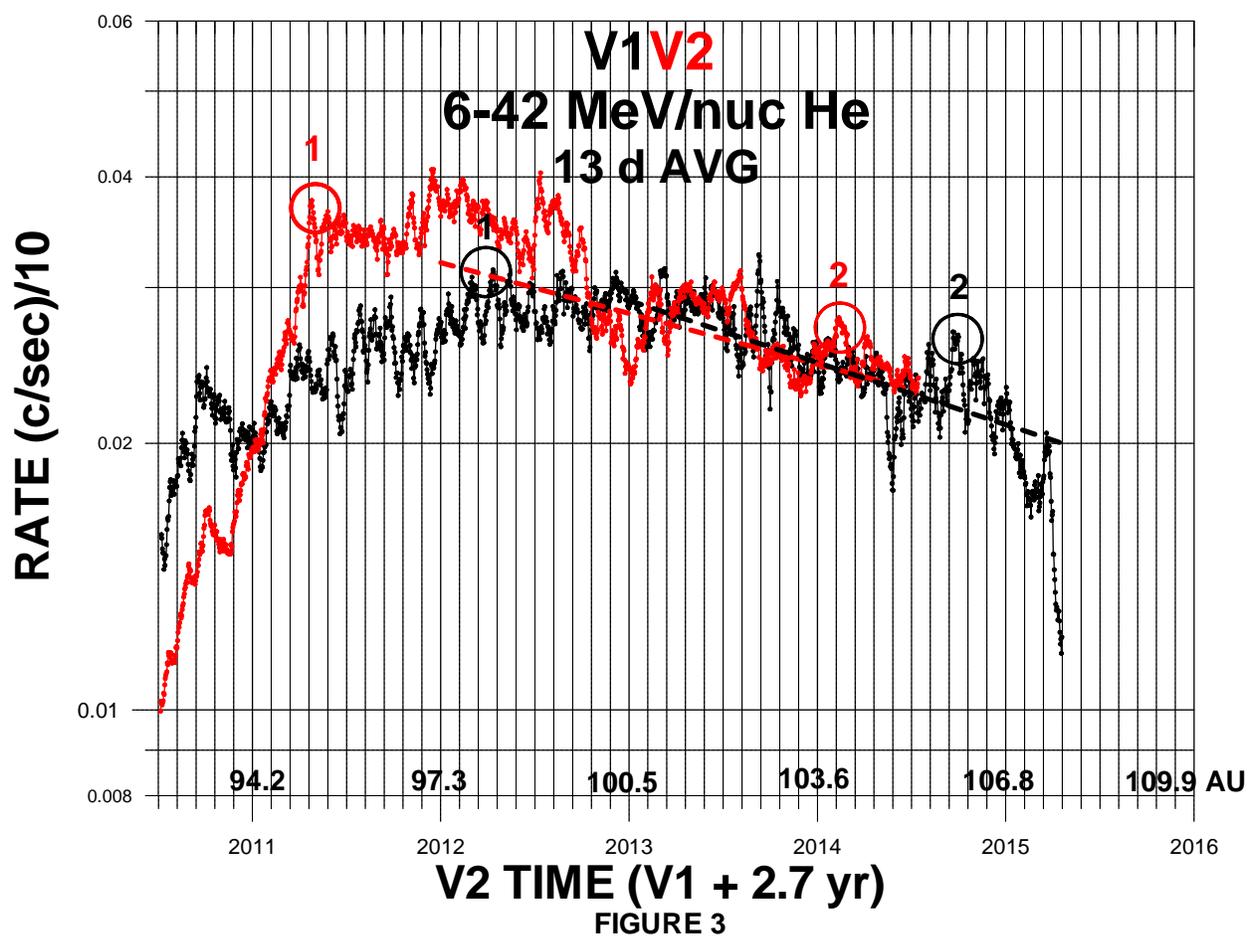

**V1 V2**
**6-42 MeV/nuc He**
**13 d AVG**

FIGURE 3